# Excitonic and lattice contributions to the charge density wave in 1$T$-TiSe$_2$ revealed by a phonon bottleneck.

H. Hedayat[1], C. J. Sayers[2], D. Bugini[1], C. Dallera[1], D. Wolverson[2], T. Batten[3], S. Karbassi[4], S. Friedemann[4], G. Cerullo[1], J. van Wezel[5], S.R. Clark[4,6], E. Carpene[7], E. Da Como[2]

1 Dipartimento di Fisica, Politecnico di Milano, 20133 Milano, Italy

2 Department of Physics, Centre Nanoscience and Nanotechnology (CNAN) and Centre for Photonics and Photonic Materials (CPPM), University of Bath, BA2 7AY Bath, UK

3 Renishaw plc., Wotton-under-Edge, GL12 7DW, UK

4 HH Wills Physics Laboratory, University of Bristol, BS8 1TL Bristol UK

5 Institute for Theoretical Physics, Institute of Physics, University of Amsterdam, 1090 GL Amsterdam, The Netherlands

6 Max Planck Institute for the Structure and Dynamics of Matter, CFEL, Hamburg, Germany

7 IFN-CNR, Dipartimento di Fisica, Politecnico di Milano, 20133 Milano, Italy

**Abstract**

**Understanding collective electronic states such as superconductivity and charge density waves is pivotal for fundamental science and applications. The layered transition metal dichalcogenide 1$T$-TiSe$_2$ hosts a unique charge density wave (CDW) phase transition whose origins are still not fully understood. Here, we present ultrafast time- and angle-resolved photoemission spectroscopy (TR-ARPES) measurements complemented by time-resolved reflectivity (TRR) which allows us to establish the contribution of excitonic and electron-phonon interactions to the CDW. We monitor the energy shift of the valence band (VB) and coupling to coherent phonons as a function of laser fluence. The VB shift, directly related to the CDW gap closure, exhibits a markedly slower recovery dynamics at fluences above $F_{th}$ = 60 μJ cm$^{-2}$. This observation coincides with a shift in the relative weight of coherently coupled phonons to higher frequency modes in time-resolved reflectivity (TRR), suggesting a phonon bottleneck. Using a rate equation model, the emergence of a high-fluence bottleneck is attributed to an abrupt reduction in coupled phonon damping and an increase in exciton dissociation rate linked to the loss of CDW superlattice phonons. Thus, our work establishes the important role of both excitonic and phononic interactions in the CDW phase transition and the advantage of combining complementary femtosecond techniques to understand the complex interactions in quantum materials.**

## I. INTRODUCTION

Charge density waves (CDWs) are an important component in phase diagrams of many correlated electron systems [1,2]. Typically observed in low-dimensional materials, the signatures of a CDW phase have been reported in two-dimensional transition metal dichalcogenides (TMDs) [3], cuprate superconductors [4], π-conjugated polymers [5] and metal oxides [6]. The central importance of CDW states arises from the relationship between fluctuations in their order parameter and superconductivity, Mott insulating states, and spin density waves [1,7]. In the TMD 1$T$-TiSe$_2$ superconductivity appears in proximity to CDW incommensurability [8], which can be achieved by pressure [9], copper doping [10], or electrostatic gating [11]. Thus, understanding of the CDW transition mechanism for this material has attracted considerable scientific interest, especially concerning its driving mechanism [12-14].

The CDW phase in 1$T$-TiSe$_2$ is achieved upon cooling below T$_{CDW}$ = 202 K, where the hexagonal lattice of the *normal phase* found at room temperature undergoes a reconstruction forming a 2a × 2a × 2c superlattice. This structural fingerprint, denoted here as *periodic lattice distortion* (PLD), occurs together with the opening of an electronic gap (Δ = 130 meV at 80 K), which is large compared to other TMDs exhibiting a CDW [2,15]. Following early experiments on TiSe$_2$, it was suggested that the CDW state is in fact an excitonic insulator stabilised by Coulomb interactions [16,17], owing to the semi-metallic character of the band structure, featuring holes in the Se-4p valence band (VB) at the $\bar{\Gamma}$–point and electrons in the Ti-3d conduction band (CB) at the $\bar{M}$-point of the first Brillouin Zone (BZ) (Fig. 1a). Thus, the presence of a PLD and excitons motivated several experimental and theoretical studies aimed at identifying the role of phonons compared to Coulomb interactions [12,15,18,19]. These efforts have highlighted 1$T$-TiSe$_2$ as a model system for studying many-body electron and phonon interactions in condensed matter physics and recently culminated with the report of Bose-Einstein condensation of excitons in this material [20].

When compared to other CDW materials of the TMD family, 1$T$-TiSe$_2$ has small atomic displacements in going from the normal phase structure to the PLD, only ~0.02 Å [21], in stark contrast to changes of up to 0.1 Å observed in 1$T$-TaS$_2$ for example [18]. The small PLD has been argued to indicate the limited importance of electron-phonon coupling in driving the CDW, thus supporting a purely excitonic mechanism [2,22]. The typical and consistently reported signatures of the CDW transition in 1$T$-TiSe$_2$ are a downwards shift of the VB energy [23], as a consequence of the CDW gap ($\Delta$) opening, and the presence of backfolded VB appearing in momentum space at $\bar{M}$, consistent with the superlattice structure as sketched in Fig. 1a (the in plane 2a × 2a reconstruction results in $\bar{M}$ being at the centre of the reconstructed BZ) [24,25]. However, these important observations, reported by steady-state ARPES, have failed to conclusively identify the excitonic or lattice contribution to the CDW.

Ultrafast spectroscopy is the experimental tool of choice to probe out-of-equilibrium phenomena in correlated electron systems [26-29]. One of the central themes in this research field is *bottleneck* dynamics, where out-of-equilibrium phonons impede excited carriers from re-joining the CDW or superconducting condensate [26,29,30]. Previous ultrafast studies on TiSe$_2$ have not reported such bottleneck effects or used it to disentangle the excitonic and electron-phonon contributions to the CDW [31-33]. Nevertheless, recent experimental evidence based on optical-pump THz-probe has shown how excitonic order can be transiently suppressed at any sample temperature below T$_{CDW}$, but with the PLD remaining robust only up to 150 K [13]. Also, signatures of phonon driven oscillations in the CDW recovery have been seen by TR-ARPES clearly suggesting a role for phonons [33]. While such reports cast doubts on a purely excitonic picture, they also open several critical questions: Are phonons and excitons only weakly coupled? How and on what time scale does the lattice dynamics contribute to the CDW recovery? More generally, to what extent is the electron-phonon coupling involved in the CDW formation? To address some of these questions which are also relevant for other CDW materials, a combination of complementary ultrafast spectroscopies is required. While most previous studies used only a single technique, a holistic understanding of the phenomena can be obtained when perturbations to the CDW induced by pump light are monitored both in the band structure, for example using ARPES, and in the PLD, by a complementary phonon sensitive spectroscopic probe.

Here, we use TR-ARPES and time resolved optical reflectivity (TRR) to clarify how phonon dynamics in 1$T$-TiSe$_2$ influence the CDW *recovery* following transient perturbation by 30 fs, 1.82 eV light pulses. In TR-ARPES, thanks to a purposely designed combination of time resolution (< 70 fs), energy resolution (~ 53 meV), and sensitivity at low laser fluence (enabled by the 80 kHz laser repetition rate), we are able to probe the VB dynamics. We find three distinct *out-of-equilibrium regimes* as a function of excitation fluence. At fluences below $F_{th}$ = 60 µJ cm$^{-2}$ the weakly perturbed CDW recovers within a short timescale of 2 ps. Above this fluence the CDW is still partially present, but its recovery exhibits a bottleneck concomitant with a change in the coherently coupled phonons seen in TRR. With the help of a rate equation model we describe how phonons contribute to the recovery dynamics. Finally, for a fluence above $F_{\text{CDW}} = 200$ µJ cm$^{-2}$, we enter a third regime with a transient complete suppression of the CDW.

## II. METHODS

High quality 1$T$-TiSe2 single crystals were grown using the chemical vapour transport method. Titanium (99.9%) and selenium (>99.9%) powders were sealed inside an evacuated quartz ampoule, together with iodine (>99.9%) which acts as the transport agent. To ensure the correct stoichiometry, a slight selenium excess was included. Single crystals with a typical size of 4 x 4 x 0.1 mm$^3$ were selected for TR-ARPES measurements. The Supplemental data, Ref. [34], shows resistivity as a function of temperature, where the salient features of the CDW transition appear below 202 K and confirms the high quality of our samples as discussed in Ref. [16].

TR-ARPES experiments were performed using a custom setup [35] based on a high repetition rate amplified Yb laser (Pharos, Light Conversion) operated at 80 kHz. The pulses from this laser, 290 fs in duration and at 1030 nm, are used to pump a non-collinear optical parametric amplifier (NOPA) which outputs 30 fs pulses at 680 nm (1.82 eV). The NOPA output is used both as pump beam in our TR-ARPES experiment and also to generate the 205 nm (6.05 eV) probe beam for photoemission, through a series of nonlinear optical processes [35]. The time resolution (cross-correlation between pump and probe pulses) is 65 fs, while energy resolution is ~53 meV. In order to measure angle-resolved photoemission we used a time-of-flight detector and spectra for different angles were recorded by rotating the crystal with respect to the analyser [34]. Our photon energy allows us to probe up to ± 0.2 Å$^{-1}$ within the BZ, which is sufficient to clearly observe the dynamics of charge carriers in the vicinity of the $\bar{\Gamma}$-point. Before TR-ARPES measurements, the 1$T$-TiSe$_2$ single crystals were cleaved in-situ to expose a clean surface, and oriented using low energy electron diffraction (LEED). Degenerate time-resolved reflectivity (TRR) experiments were performed with the 30 fs pulses at 680 nm as pump and probe beams. They impinged on the sample surface at about 45° with crossed polarisations in order to avoid interference artefacts. All the experiments were performed at a sample temperature ranging from 80 K to 300 K as specified in the figures.

## III. EXPERIMENTAL RESULTS

In TR-ARPES, the infrared pump pulse first promotes electrons from the occupied to the unoccupied states with the same momenta, while the subsequent UV probe pulse is used to photoemit electrons and the transient energy dispersion is mapped in momentum space. Figures 1(b)-1(d) show the evolution of the TR-ARPES maps at three different time delays between the pump pulse and the probe (6.05 eV) for a TiSe$_2$ single crystal at 80 K. The sample temperature is ideal since it is below T$_{CDW}$ = 202 K, but sufficiently high to allow perturbations to the PLD as we show below. Figure 1(b) shows the TR-ARPES map at -225 fs delay (i.e. before the pump photoexcitation), which reflects the VB dispersion in the vicinity of the $\bar{\Gamma}$ point ($k_\parallel = 0$) in the CDW phase along the $\bar{K}-\bar{\Gamma}-\bar{K}$ direction. The effect of the pump is apparent in Fig. 1(c) at time delay +25 fs where electrons from the VB have been promoted into a high energy CB for $k$-states at the edges of our detection window. As a high energy CB becomes transiently populated, photoelectron signal is observed above the Fermi level. The effect of the pump and the TR-ARPES maps exhibit features similar to what has been reported in other TR-ARPES studies [32,36,37]. Like in other semimetals [38,39], carrier relaxation from high energy states occurs within a few hundred fs (see Appendix A).

The key signature of the CDW is the gap, $\Delta$: Fig 1(d) shows that the VB is shifted upwards in energy, i.e. $\Delta$ is reduced, even at +425 fs delay. This VB shift lasting longer than the pump laser duration (30 fs) is a signature of laser-induced perturbation of $\Delta$. Changes in the VB binding energy are extensively documented in *steady-state* ARPES, when heating the 1$T$-TiSe$_2$ lattice from T < T$_{CDW}$ to the normal phase [15,23]. In order to accurately study the VB energy shift, we have performed an analysis at the fixed detection angle -14º ($k_\parallel \approx$ - 0.1 Å$^{-1}$ for the VB) (dotted white line in Fig.1 b-d) as a function of pump-probe time delay. This angle allows better separation between the Fermi level and the VB, thus avoiding effects due to thermal smearing of the Fermi-Dirac distribution.

As confirmation that the energy shift is indeed a dynamic perturbation of the CDW, and not just heating of the VB electron distribution, we show in Figs. 1(e) and 1(f) the VB dynamics at 80 K and 300 K, respectively. While the data at 80 K exhibit a clear upshift in the VB energy lasting for the whole measurement window of 3 ps, the 300 K trace shows no energy shift and only a loss of intensity due to the transient depletion of electrons in the VB. A closer examination of the VB photo emission spectrum (PES) allows us to identify an energy position that shows no dynamics upon pumping at 300 K, i.e. a nodal point in the energy distribution curves that occurs at about 30% of the maximum intensity, see Appendix B. This intensity value is used hereafter to monitor the VB dynamics.

The possibility to tune the pump fluence gives additional important insights. Figure 2 illustrates how the VB shift and spectral weight are influenced. A crucial finding of our study is the measurement of the fluence, $F$, necessary for a transient closure of $\Delta$, reported to be $\Delta = 0.13$ eV at 80 K [23]. For the analysis, it is important to note that the dominant contribution to the opening of $\Delta$ comes from a VB downshift of 0.11 eV at $\bar{\Gamma}$, observed upon cooling from room temperature to 80 K [23]. Figure 2(a) shows the time dependence of the VB shift for selected pump fluences. The *maximum* VB shift occurs after about 200 fs in all traces, and for 250 μJ cm$^{-2}$ reaches ~0.1 eV. The rise time in the VB shift occurs on time scales almost identical to the folded VB ARPES intensity suppression at $\bar{M}$, recently measured in a comparable fluence range by Buss et al. [40]. Figure 2(b) reports the VB shift as a function of fluence at specific time delays corresponding to the maximum shift (dots) and at $t = 3$ ps (triangles), the latter will be discussed below. To gauge the VB maximum shift with respect to the level of CDW perturbation, we have also plotted the shift in the equilibrium VB binding energy as horizontal lines in Fig. 2(b), taken from high resolution steady-state ARPES [23] in going from 70 K to 300 K. Two key observations are apparent from the trend of the VB maximum. First, the maximum shift is initially linear for low fluence before reaching a plateau for $F > 93$ μJ cm$^{-2}$ equivalent to a shift in the equilibrium VB binding energy observed at temperatures between 180 K and 200 K. Second, this plateau persists until a critical fluence of $F_{\text{CDW}} = 200$ μJ cm$^{-2}$, beyond which the VB transiently shifts above the 200 K line of equilibrium data and is consistent with complete suppression of $\Delta$ and disappearance of CDW order. These nonlinear trends are not due to saturation of absorption from TiSe$_2$, Appendix A, or average laser heating [34], but are instead an intrinsic characteristic of the CDW dynamics.

Returning to the VB dynamics in Fig. 2(a) for the low fluences of 31 and 62 μJ cm$^{-2}$ we find a fast recovery, described by a mono-exponential decay with equal time constants of ~ 770 fs. This leads to a complete VB recovery, i.e. 0 eV shift, at time delays > 2 ps. For higher fluences, multi-exponential decays with lifetimes longer than 1100 fs are found. The inset in Fig. 2(a) gives a clearer comparison of the same data and plotted on a logarithmic normalised intensity scale. The dynamics can be grouped into two well-defined categories and points to a threshold fluence, $F_{\text{th}}$, between these regimes of fast and slow VB recovery. The residual VB shift at 3 ps (triangles), shown in Fig. 2(b), clearly identifies $F_{th} > 62$ µJ cm$^{-2}$. Note that $F_{th} \cong F_{CDW}/3.3$, and so does not coincide with the complete suppression of $\Delta$ occurring at $F_{CDW}$.

Following these observations, we look at the spectral *intensity* dynamics from the TR-ARPES. The left panel in Fig. 2(c) shows the spectral weight obtained by integrating the PES up to the Fermi level for $F < F_{th}$, normalised as described in caption. For both fluences the spectral weight is depleted upon photoexcitation and re-established within 500 fs. However, for the 62 µJ cm$^{-2}$ data (orange curve) at a delay > 500 fs the spectral weight shows a small and short-lived intensity *gain*. Similar behaviour but more pronounced and longer lasting, is observed for the two higher fluences reported in the right panel of Fig. 2(c). Photoexcitation with $F \geq F_{th}$, therefore increases spectral weight in the VB. To confirm the origin of this gain we compare the traces at 80 K and 300 K in Fig. 2(d). At negative delay, the total intensity measures electrons in the VB up to the Fermi level. This is diminished when going from 300 K to 80 K because CDW formation transfers spectral weight from $\bar{\Gamma}$ to the VB folded at $\bar{M}$, c.f. Fig. 1(a). Crucially, the 300 K data in the normal phase do not show any increase in intensity above the initial value. Thus, the observed gain (blue arrow) is caused by the photo-induced unfolding of the VB from $\bar{M}$ to $\bar{\Gamma}$ points and indicates breaking of exciton pairs in the CDW condensate [36], and/or a disturbance of the PLD.

The results of Fig. 2 report the 1$T$-TiSe$_2$ VB dynamics in a fluence regime rarely studied by TR-ARPES [40]. It is important to point out that previous time resolved all-optical experiments in the low fluence regime $F < F_{th}$ have clarified how the initial perturbation of the CDW by fs laser pulses is non-thermal, i.e. it concerns mainly the electronic order in the system represented by the exciton condensate and to a negligible extent the lattice degrees of freedom [13,14]. This is consistent with electron-electron and electron-exciton scattering times on the order of hundreds of femtoseconds [13]. When discussing possible scenarios for the CDW in TiSe$_2$, we consider that the CDW formation is due to both excitonic *and* lattice interactions, where the relative contributions and relationship are currently unknown. For fluences below $F_{th}$, the rapid VB recovery is consistent with electronic dynamics and suggests that mainly the excitonic part of the CDW is perturbed. Above $F_{th}$ there is an additional contribution with a longer recovery time, indicating a bottleneck in re-establishing the CDW ground state.

Interestingly, $F_{th}$ identified from the 3 ps data in Fig. 2(b) corresponds to a VB position at 150 K from steady-state measurements. This is the sample temperature at which recent femtosecond THz experiments reported the disappearance of the phonon fingerprint of the PLD [13]. All together we suggests that the < 200 fs non-thermal shift of ~0.05 eV at $F_{th}$ can be interpreted as a lower limit for the initial excitonic contribution to the CDW. Above $F_{th}$, a second weakening process for Δ plays a role. Notice that a simple picture in which excitonic and phononic contributions can be obtained separately from the VB dynamics alone, and summed to obtain the full Δ, does not apply. Rather, we now look at how phonons influence the dynamics of Δ.

Understanding of the CDW bottleneck dynamics benefits from monitoring lattice degrees of freedom. While steady-state ARPES experiments can provide indirect information on lattice structure from changes in the electronic band structure, time-resolved experiments allows the dynamics of a subset of phonons to be probed in real time [41]. In TR-ARPES, we observe that below $F_{th}$ the VB dynamics are modulated by periodic oscillations, very likely connected to coherent phonons, whereas above $F_{th}$ their amplitude weakens or is undetectable [34]. Oscillations of the VB in TR-ARPES signify the presence of phonons connected with the order parameter, Δ thus it is conceivable that such oscillations will weaken as the CDW is perturbed above $F_{th}$. Further information on phonon dynamics can be obtained from optical TRR experiments which offer a slightly higher time resolution and probe the change in refractive index of our crystal modulated by phonons. Fig. 3(a) shows clear oscillations in TRR for all fluences. It is important to note that below $F_{th}$, both the period and damping time is similar to those seen in TR-ARPES. Figure 3(b) illustrates examples of the oscillatory component of the signal at different pump fluences after subtraction of an exponential decay. A Fourier transform (FT) of the data from Fig. 3(b) allows us to identify two well separated oscillation frequencies: a low frequency mode at 3.36 THz (112 cm$^{-1}$) and a high frequency mode at 6.03 THz (201 cm$^{-1}$). The lower frequency is that of the Raman active $A_{1g}$* phonon of 1$T$-TiSe$_2$ [34,42]. At $F \leq F_{th}$ the $A_{1g}$* is the most intense coherently coupled phonon with the largest amplitude. This mode is a consequence of the PLD and is not present in the normal phase of 1$T$-TiSe$_2$ [42]. We show in Ref. [42] that the $A_{1g}$* mode is selectively coupled to the CDW and recent experiments by Monney et al. have shown how it modulates the CDW recovery after quasi-resonant excitation at Δ. At $F \geq 132$ μJ cm$^{-2}$ the oscillations are instead dominated by the higher-frequency mode, similar in frequency to the $A_{1g}$ phonon of the normal phase [34,42], or two zone-edge modes triggered by a second order process [43,44]. Both assignments for the high frequency 6.03 THz oscillations are consistent with a perturbed PLD and excitation of phonon modes of the normal phase structure [34].

The progressive change of amplitude between the selectively coupled phonons (SCP) as the laser intensity increases can be interpreted with two hypotheses: (i) $A_{1g}*$ phonons that do not couple coherently in the CDW recovery, (ii) a loss of PLD (disappearance of $A_{1g}*$) and thus rearrangement of the lattice towards the normal phase, as also supported by the unfolding results, Fig. 2(c). Information on phonons is relevant for the bottleneck, since a substantial population of excited vibrational modes (hot phonons) can transfer energy back into the already perturbed exciton condensate and suppress its re-establishment. Most important is the observation that phonons linked to the normal phase structure modulate the dynamics when the bottleneck in the recovery appears, as clearly shown by the comparison of FT amplitudes and VB shift in Fig. 3(d).

## IV.    MODEL AND DISCUSSION

We have performed a series of simulations capable of describing the VB dynamics as a function of laser fluence. These are inspired by the Rothwarf-Taylor model which was initially used to describe the equilibration of Cooper pairs in superconductors with hot electrons and phonons [26,45] and has also been applied to CDW materials [46]. Photoexcitation dynamics in 1$T$-TiSe$_2$ is tracked via three populations: hot carriers (electrons and holes) $n_e$ created by the pump; low energy unbound quasiparticles (QPs), $n_q$, originating from breaking excitons into electrons and holes (a process consistent with Fig. 2(c)), and a population of SCPs, $N_p$. Following the pump pulse, carriers rapidly relax via electron-electron and electron-phonon scattering, breaking excitons into unbound QPs and exciting SCPs and other phonons in the process. Figure 4(a) outlines the main processes resulting in the relaxation bottleneck and a scheme of all the physical processes accounted for in the model is in Fig. S10 [34]. QPs can recombine into the exciton condensate with a rate $R$ by exciting the SCPs further, while absorption of SCPs by the exciton condensate can dissociate excitons to create QPs with a rate $\eta$. The SCPs also equilibrate with the bath through anharmonic decay at a rate $\gamma$. The complete model is discussed in Supplementary [34].

Figure 4(b) shows that the VB dynamics can be accurately modelled approximating the VB shift as $\Delta(t) - \Delta_{80K}$ via [26]

$$\Delta(t) = \Delta_{80K}\sqrt{1 - n_q(t)/n_c}, \qquad (1)$$

where $\Delta_{80K}$ = 130 meV is the CDW gap at 80 K, ref.[23], and $n_c$ is the critical QP density, which we have estimated based on our Hall effect measurements [34] and reports from literature [19].

In the model $n_e$ has been determined from the measured laser pulse fluences. The fluence dependence of the two main fitting parameters $\gamma$ and $\eta$, reported in Fig. 4(c), shows a correspondence with the VB shift behaviour and to the FT amplitudes of the different coherent phonons in Fig. 3(d). The bottleneck effect emerges in the model due to the SCPs becoming simultaneously less damped (smaller $\gamma$) and scattering more frequently with excitons (larger $\eta$). The model does not distinguish between $A_{1g}*$ and $A_{1g}$ phonons. Thus the substantial change in $\eta$ at $F_{th}$ implicitly accounts for the switch between these modes seen in TRR experiments, and for the likely change in exciton-phonon scattering rates in the weakened PLD. A correlation between the values of $\gamma^{-1}$ and the damping time of coherent oscillations $\tau_{damp}$ extracted from TRR experiments appears from Fig. 4(c). This quantitative correspondence for the same physical observable, inferred from the modelled VB dynamics and independently extracted from TRR data, lends weight to the credibility of our model. Future work with structural probes such as TR x-ray diffraction will help to probe the anharmonic decay of SCP in momentum space [43,47] and thus further verify such correlation. The ratio $\eta/\gamma$ reported in Fig. 4(d) is an important metric for which the strong bottleneck regime occurs when $\eta/\gamma < 1$ [26]. In contrast to superconductors, where the bottleneck regime arises directly from the nonlinearity of the R-T equations without changing $\eta$ and $\gamma$ as a function of fluence, in CDW materials the electronic order is coupled to the PLD. Weakening of the PLD results in dramatic changes to the interactions at play in the recovery of excitons as illustrated by Fig. 4(d).

A comparison with previous femtosecond experiments performed at lower sample temperatures supports our conclusions that the anharmonic decay of SCP plays a role in the CDW bottleneck dynamics [13,33]. Monney et al. observed coherent $A_{1g}*$ phonon oscillations in the recovery of the folded band intensity at the $\overline{M}$ point upon excitation with pulses at 400 meV photon energy and a fluence of 400 μJ cm$^{-2}$ [33]. While this laser fluence is almost an order of magnitude larger than our $F_{th}$, we note that the sample temperature was 30 K and the photon energy is comparable to the CDW gap, thus generating directly quasiparticles with the pump pulse instead of hot free carriers. A pump photon energy of 1.55 eV, i.e. more similar to the one used by us, was employed in the broad-band THz spectroscopy experiments of Porer et al. [13], where a delayed recovery of the excitonic order was seen at fluences above 40 μJ cm$^{-2}$, i.e. comparable to our $F_{th}$, and with an additional suppression for higher fluences. From our data, the additional suppression can be related to phononic effects. Moreover, in agreement with our arguments that the anharmonic decay influences the CDW recovery, Porer et al. observed that the sample temperature had to be increased to 150 K in order to show a photo-induced melting of the PLD. Combining these results, it is evident that the sample temperature is an important factor. The temperature of the heat bath can control $\gamma$, leading to much longer damping times as temperature is decreased below 60 K as confirmed in the TRR data of ref. [14].

We conclude that below $F_{th}$ only few QPs are excited out of the excitonic condensate and relax by coupling to $A_{1g}*$ phonons characteristic of the PLD. Above $F_{th}$ the excitonic part of the CDW gap is perturbed to an extent where the coupling to the $A_{1g}*$ phonons is weakened and the QP population is substantial. Thus, recovery of the CDW experiences a bottleneck controlled by the anharmonic decay of hot phonons and the modified phonon dispersion of the weakened PLD. Our results therefore provide evidence that in the out-of-equilibrium phases probed in our experiments, the reformation of the CDW in 1$T$-TiSe$_2$ upon electronic cooling is always influenced by lattice degrees of freedom. In addressing the introductory questions on exciton-phonon coupling, we come to the following conclusions: i) we exclude that 1$T$-TiSe$_2$ should be classified as typical Peierls CDW material, since the initial perturbation of $\Delta$ occurs on a short time scale ~200 fs consistent with electronic screening of Coulomb interactions (Fig.2a), in agreement with similar conclusions made by others [31,32,40], ii) for the quasiparticles, i.e. electron and holes at band-edge, the selective coupling to $A_{1g}*$ phonons of the PLD is involved in the exciton condensation and thus a selective coupling with specific modes should be considered strong, iii) ultimately, if PLD phonons are out of thermal equilibrium with the lattice bath, or not present at all (due to a loss of PLD), $\Delta$ is recovered only to its corresponding steady-state value of ~150 K.

We argue that these low fluence regime data contribute to the discussion on the mechanism of CDW formation in 1$T$-TiSe$_2$ in the absence of light excitation, i.e. in the ground state. The choice of looking at the recovery dynamics allows us to extract relevant parameters in understanding how SCPs of the PLD structure and of the normal structure interact with charge carriers and excitons. Most importantly our work highlights the role of anharmonic coupling between SCPs and the thermal bath, which in our simplified model is the lattice temperature set by other low energy phonon modes. As shown by Weber et al. optical phonon modes interacting with an acoustic branch experience a softening as temperature is lowered close to $T_{CDW}$ [48]. Our work probes how phonons are directly influencing $\Delta$ and provides estimates for the scattering rates between excitons and SCPs.

To generalize the discussion and extend it to other TMD materials exhibiting a CDW transition, it is interesting to compare our results with the very recent work of Shi et al. on 1$T$-TaSe$_2$ [49]. The CDW transition temperature in 1$T$-TaSe$_2$, 470 K, is more than twice that of TiSe$_2$ and the system is known to exhibit laser induced perturbation of the electronic order at much higher fluences than those we report here [31,50]. Interestingly, Shi et al. show a transient energy shift of CDW-related Tantalum bands as a function of fluence by TR-ARPES. This is consistent with what we show in Fig. 2(b), although on a completely different scale of fluences. In 1$T$-TaSe$_2$, even after the pump induced band shift has reached a plateau, the coherent phonon oscillations related to the PLD are seen, uncovering a new metastable state mediated by selective electron-phonon coupling [49]. What is remarkable, is that the damping time of those oscillations shows similarities to our observations of Fig. 3(c), i.e. the damping time is faster as the electronic gap is closing. Such behaviour could signal the fact that a phonon bottleneck is involved in the formation of the new metastable state in 1$T$-TaSe$_2$ as well. As commented by Shi et al. this could be explained as a modulation of the electron-phonon coupling, a scenario in line with the Peierls-Mott nature of the CDW in 1$T$-TaSe$_2$ [50]. The different nature of the CDW in TiSe$_2$, where the excitonic and phononic contributions cooperate to generate a gap, is seen in the switching of the coupled phonon modes as the CDW gap closes. Our combined TRR/TR-ARPES approach could be used to determine at which fluence the PLD of 1$T$-TaSe$_2$ is perturbed to an extent where phonons of the normal phase play a role. It may also be applied to other CDW TMDs, since many of them exhibit detectable changes in PLD through Raman resonances across the respective T$_{CDW}$ [51].

## V. CONCLUSIONS

In summary, the CDW dynamics of $1T$-TiSe$_2$ following ultrafast photoexcitation shows different out-of-equilibrium regimes accessed by changing fluence. There exist two distinct regimes below the complete melting of the CDW, one in which the PLD structural order appears robust and a weak perturbation of the excitonic condensate results in a fast recovery time of < 2 ps. For fluences above 60 μJ cm$^{-2}$ the CDW is still present in a metastable state controlled by the lattice degrees of freedom, i.e. hot phonons and partial loss of PLD. Our study conclusively shows that the dynamics of Δ can be modelled as the increase of QPs excited out of the exciton condensate, which in turn depend on SCPs. In the absence of full thermalization of this subset of phonons with the bath, the full recovery of the CDW is impeded. Future work aiming at controlling the CDW by targeted excitation of phonons with intense THz pulses could represent a new avenue to connect the incommensurate CDW in 1$T$-TiSe$_2$ to the superconductivity [11] as well as three pulse experiments to investigate the symmetry of the weakened PLD [52]. The properly designed combination of ultrafast spectroscopy methods presented here, TRR and TR-ARPES, has proven decisive in giving insights into the complex nature of the CDW in $1T$-TiSe$_2$. This experimental approach can be applied to many other correlated electron materials such as superconductors and topological insulators in order to disentangle the complex many-body interactions that dominate their unique properties.

## ACKNOWLEDGEMENTS

We thank P. Jones, P. Reddish and W. Lambson for technical support. C.S. and S.K. acknowledges funding and support from the Engineering and Physical Sciences Research Council (EPSRC) Centre for Doctoral Training in Condensed Matter Physics (CDT-CMP), Grant No. EP/L015544/1. S.R.C. and D. W. acknowledge support from EPSRC under grants Nos. EP/P025110/1 and EP/M022188/1, respectively. J.v.W. acknowledges support from a VIDI grant financed by the Netherlands Organization for Scientific Research (NWO). H.H. acknowledges financial support through the Postdoctoral International Fellowship program of Politecnico Milano. S.F. acknowledges support from EPSRC under grant No. EP/N026691/1. E.D.C. and J.v.W. wish to thank the Royal Society for a Research Grant and an International Exchange Program. E.D.C. acknowledges support from Horizon 2020 (654148, Laserlab-Europe). G.C. acknowledges support by the European Union's Horizon 2020 research and innovation programme under grant agreement No 785219 (GrapheneCore2).

**Corresponding author**

Correspondence to Enrico Da Como edc25@bath.ac.uk

## APPENDIX A: TIME-DEPENDENCE OF FREE CARRIER POPULATION AND LINEARITY WITH PUMP FLUENCE

In order to check the linearity of the laser pumping effect in our TR-ARPES experiment, we have analysed the total intensity above the Fermi level, E$_F$, for all spectra, which provides an indication of the transient free carrier population induced by the pump pulse. For this, we used the normalised spectra in Fig. 6 at each pump-probe delay and integrated across the high energy tail to intensity $\leq 0.25$ (just below the nodal point of all spectra in the normal phase) as shown by the example in the inset of Fig. 5(a). The main panel of Fig. 5(a) shows the result of the integration as a function of pump-probe delay for each fluence in the CDW phase (solid lines) and in the normal phase (dashed line) for comparison. All curves have a maximum near + 50 fs and recover to their value before pump excitation (at -225 fs) within 1500 fs. Fig. 5(b) shows the integrated signal at negative (-175 fs) and at positive (+ 50 fs) time delays for different fluences. Importantly, the (-175 fs) tail contribution, black dots in Fig. 5(b), is found to be the same at all fluences within the experimental error. This is consistent with the tail being an equilibrium effect. If we subtract the tail at negative times, represented by the black dots, then all data points scale down, giving an intercept passing through the origin for the line of best fit. The y-axis error of the integrated spectral intensity is related to the statistical error ($\sqrt{N}$) in the number of counts, N and the x-axis error is the fluctuation of laser fluence estimated with an optometer and by analysing the fluctuation of the VB peak intensity measured at positive delays multiple times under the same conditions with high statistics. Further aspects on the linearity of photoexcited carrier population can be found in Ref. [34].

**APPENDIX B**: **KEY SIGNATURES OF CDW SUPPRESION IN TR-ARPES**

In order to highlight all the spectroscopic features of CDW suppression seen in our experiment, the photo emission spectra (PES) for various pump-probe delays are shown in Fig. 6, these were extracted from the data in Fig. 1(e – f) of the main text. In the normal phase at 300 K, Fig.6 (a), shortly after the pump pulse (+ 125 fs), there is a clear reduction of the valence band intensity and a significant broadening of the photoemission peak across the Fermi level. We do not observe any shift in the VB energy (horizontal black arrows) and the intensity at long delay (+2425 fs) returns to the initial value (at -225 fs). At sample temperatures below $T_{CDW}$ the same initial reduction in intensity and broadening of states occurs (+125 fs, Fig.6 (b)). However, in contrast to the normal phase data, there is a clear upshift of the VB energy (horizontal red arrows). This up shift is a signature of the photo-induced CDW gap closing. In addition, at long delay ($\geq$ 1ps) there is an *increase* in the VB intensity (blue shaded region) which we attribute to the unfolding of the VB replica from $\bar{M}$ to $\bar{\Gamma}$ as the CDW order is suppressed as discussed in Fig. 2(c-d).

**Figures and captions**

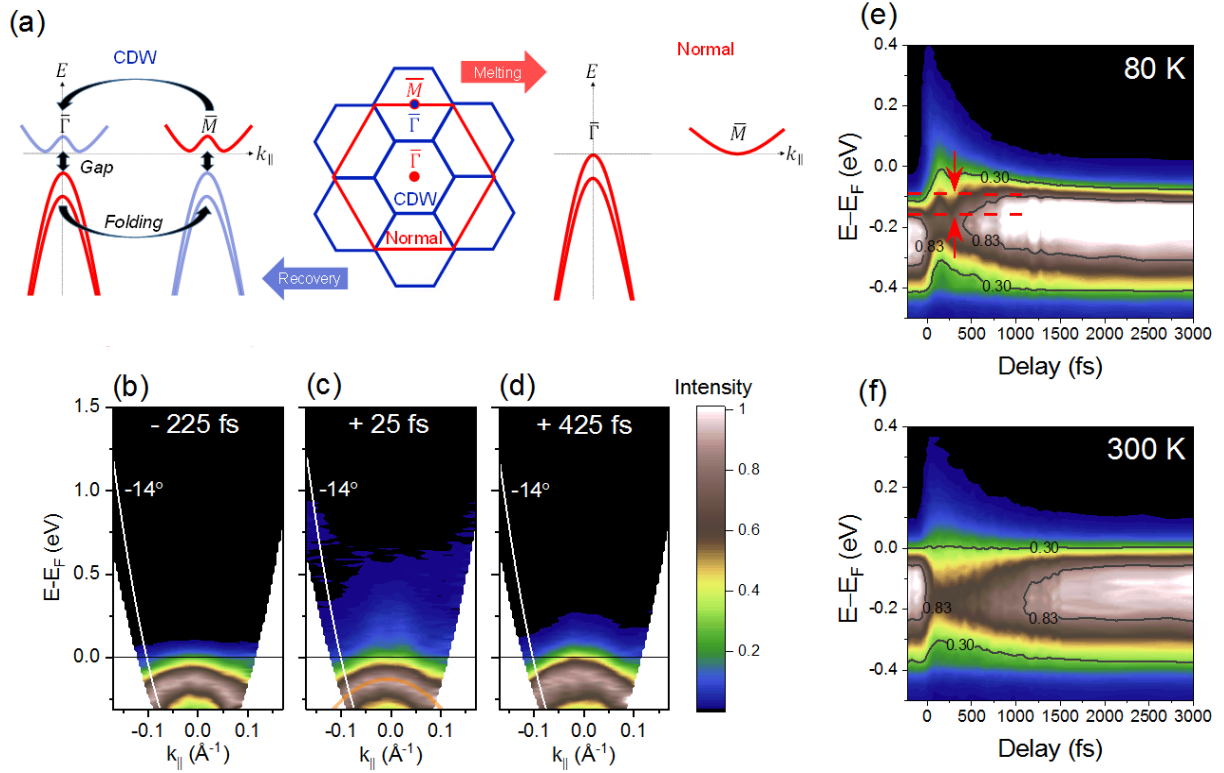

**Fig. 1.** Valence band dynamics in the CDW and normal phase of 1*T*-TiSe$_2$. (a) Sketch of the TiSe$_2$ band structure when switching between the CDW phase (blue) and normal phase (red), together with the projection of the first Brillouin zone (BZ) similar to the one reported in ref. [53]. Band folding arising from the 2a x 2a x 2c PLD is indicated by the curved arrows as $\bar{\Gamma}$ and $\bar{M}$ become equivalent. Vertical arrows show the lowering of the VB maximum due to the formation of a CDW gap, Δ near the Fermi level. (b-d), Time evolution of the ARPES maps in the CDW phase along the $\bar{K}$–$\bar{\Gamma}$–$\bar{K}$ direction at select pump-probe delays for 125 μJ cm$^{-2}$ fluence. The orange line in (c) is a guide to the eye for the VB dispersion. It corresponds to an effective mass of -0.35m$_e$ which is comparable to previous reports [23,54]. The -14° angle for the analysis of the VB dynamics is shown as the white solid line. (e-f), VB dynamics at 80 K and 300 K, respectively, for a pump fluence of 250 μJ cm$^{-2}$. The red dashed lines in the 80 K plot indicate the maximum VB shift after pump excitation. The black curves indicate contours at different ARPES intensity. Close inspection of the VB spectra (panel e-f) reveal two subbands which are analysed separately in the Supplementary Information for completeness [34].

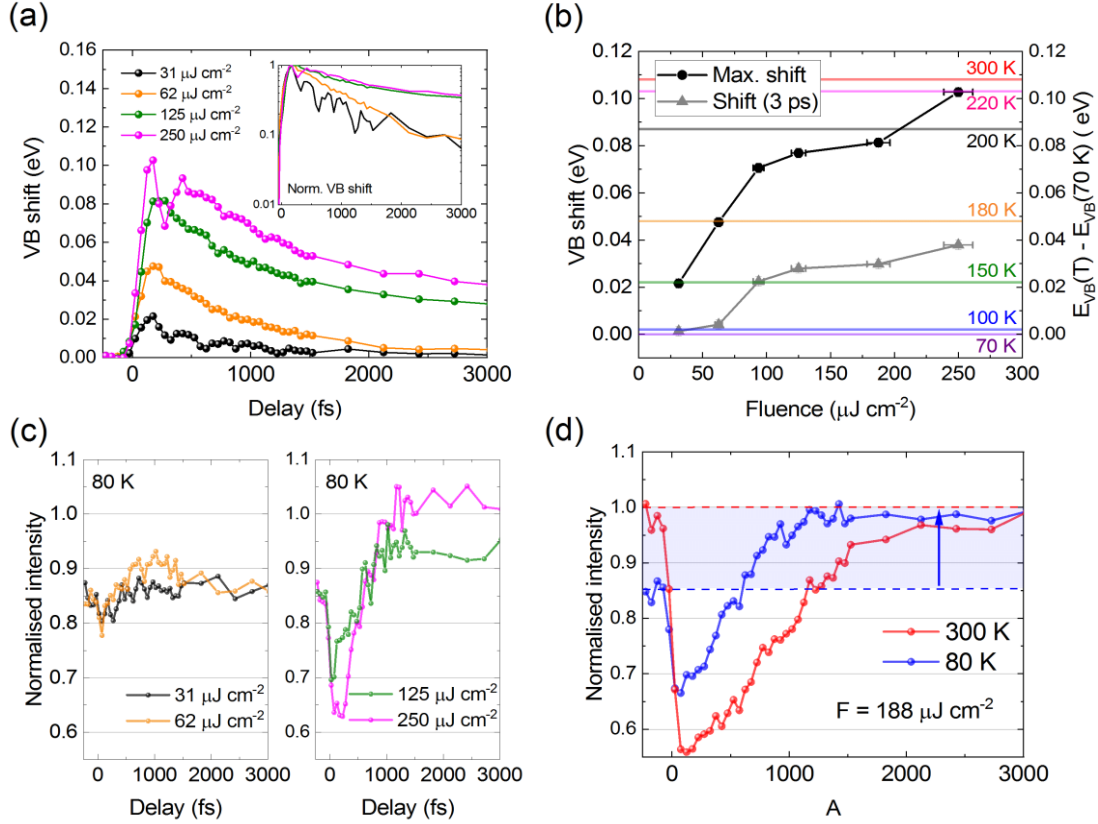

**Fig. 2.** Fluence dependence of the VB dynamics and spectral intensity. (a), VB energy shift, referred to the unperturbed position, as a function of pump-probe time delay at different pump laser fluences as indicated in the main panel. VB position has been extracted from the contour at 0.3 of the maximum ARPES intensity (example in Fig 1e). The inset shows the same data normalised to the maximum VB energy shift on a logarithmic scale. (b), Maximum VB shift (dots) and shift at 3 ps (triangles) extracted from panel (a) as a function of fluence. The horizontal lines are linked to the right y-axis and are the shift in VB energy determined by high resolution steady state ARPES as the sample temperature is increased from 70 K, adapted from reference [23]. The VB position at 70 K from steady state ARPES has been set to coincide with the VB energy at negative delays in our experiments. Energy error bars on VB shift data points are < 2 meV. (c), Intensity of the VB at 80 K and for different laser fluences as indicated, normalised to ARPES intensity at negative delays and at room temperature (shown in d). (d), Comparison between the normalised VB intensity in the CDW phase (80 K) and normal phase (300 K). A gain in intensity is observed in the 80 K data indicated by the blue shaded region.

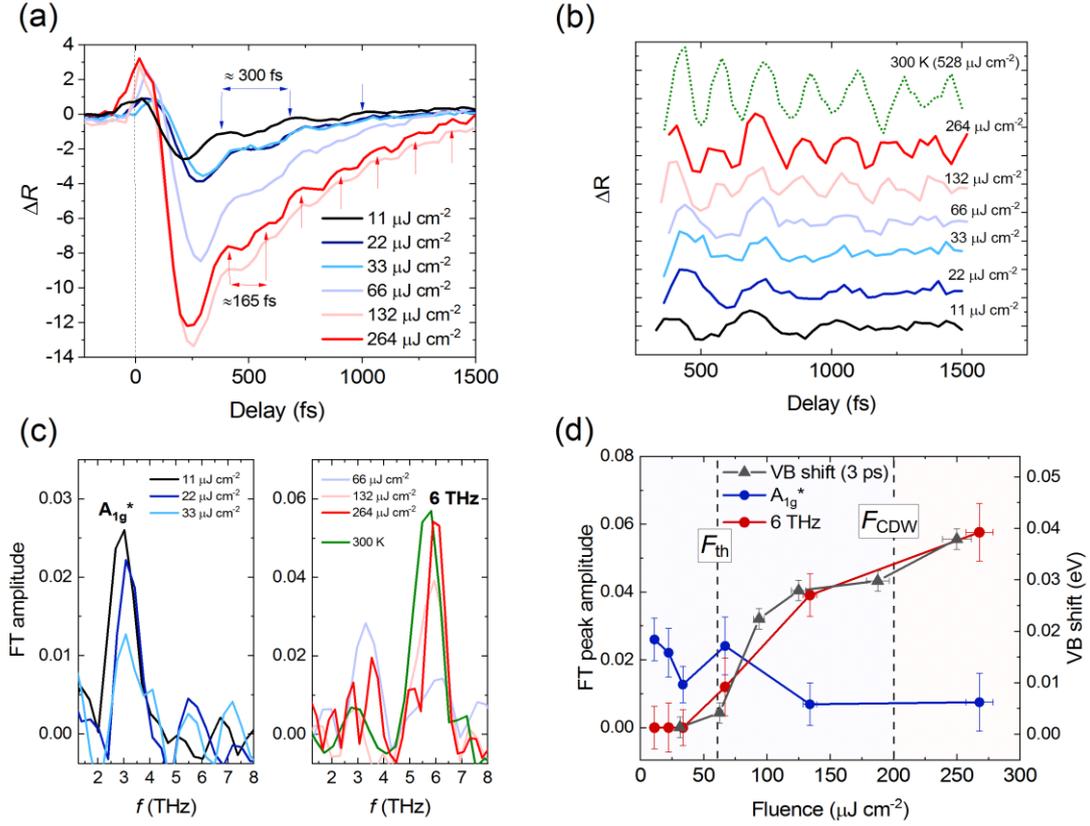

**Fig. 3.** Coherent coupling to phonons in TRR. (a) Coherent phonon oscillations observed by time-resolved reflectivity (TRR) in the CDW phase (80K) at different laser fluences. (b), Oscillations after subtraction of an exponential decay in the TRR data in panel (a). (c), Fourier transform (FT) amplitude for the oscillations in panel b at 80 K together with the normal phase (300 K) for comparison. (d), FT amplitude of the $A_{1g}*$ and 6 THz modes, as a function of fluence. The VB shift at 3 ps (black triangles) from Fig. 2b is superimposed (right y-axis).

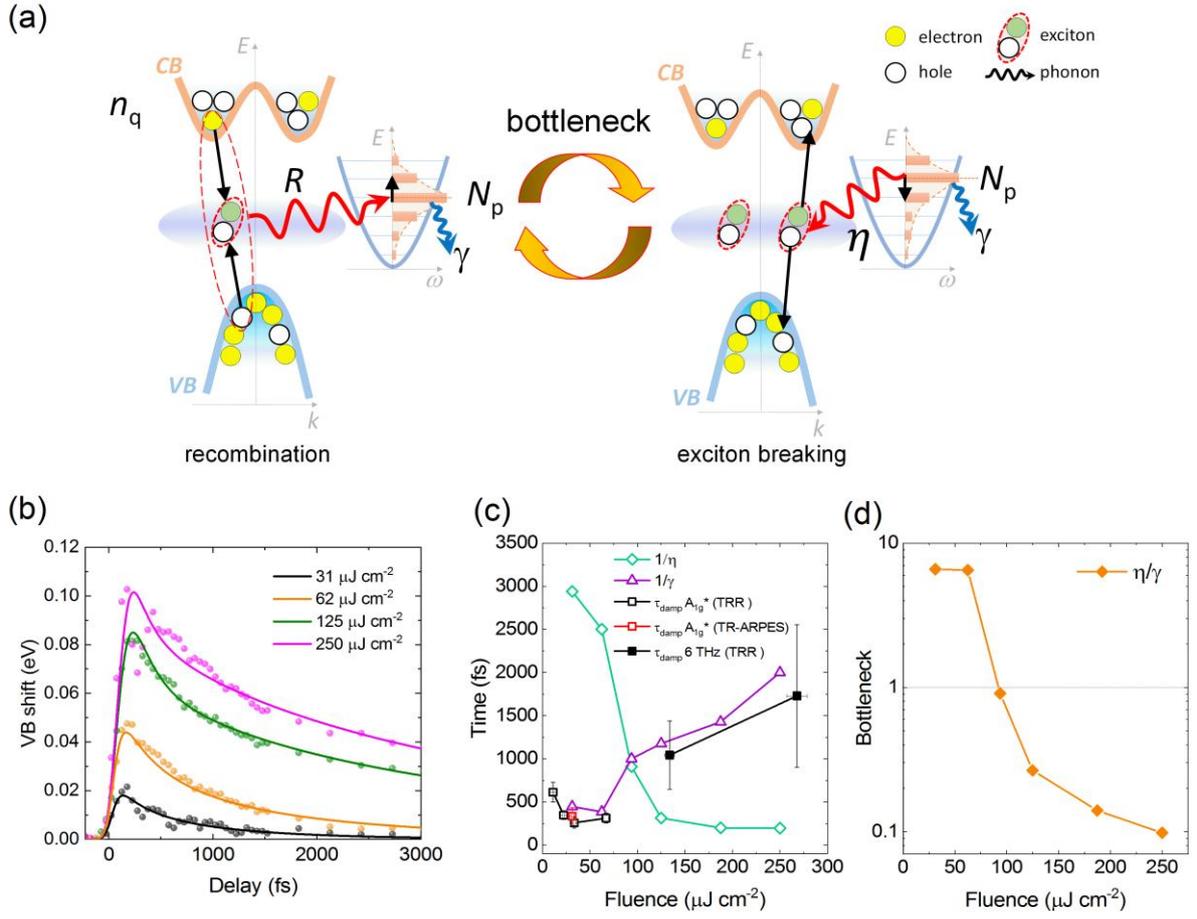

**Fig. 4.** Modelling of VB dynamics and phonon bottleneck in the CDW recovery. (a), Bottleneck involving the quasiparticle population, $n_q$, and the out of equilibrium SCPs, $N_p$. The straight black arrows indicate the transition from electron and hole quasiparticles to the exciton condensate in both directions. Red curved arrows indicate the scattering of two quasiparticles with an SCP in the recombination process (left panel) or the scattering of an exciton with an SCP generating two quasiparticles (right). γ is used to indicate the anharmonic phonon-phonon scattering with the thermal bath. The orange arrows illustrate the iterative dynamics between recombination and re-excitation influenced by γ. The Supplementary (Fig. S9) includes a further diagram illustrating the fast photoexcitation process preceding this bottleneck dynamics [34]. (b), VB dynamics extracted from the model (solid curves) overlaid with the TR-ARPES data (dots) from Fig. 2(a). (c), Fluence dependence of the fitting parameters $1/\eta$ (green diamonds) and $1/\gamma$ (purple triangles) together with damping times, $\tau_{damp}$, from experiments as indicated in the legend. (d) Ratio $\eta/\gamma$ as a function of pump laser fluence extracted from the model simulations.

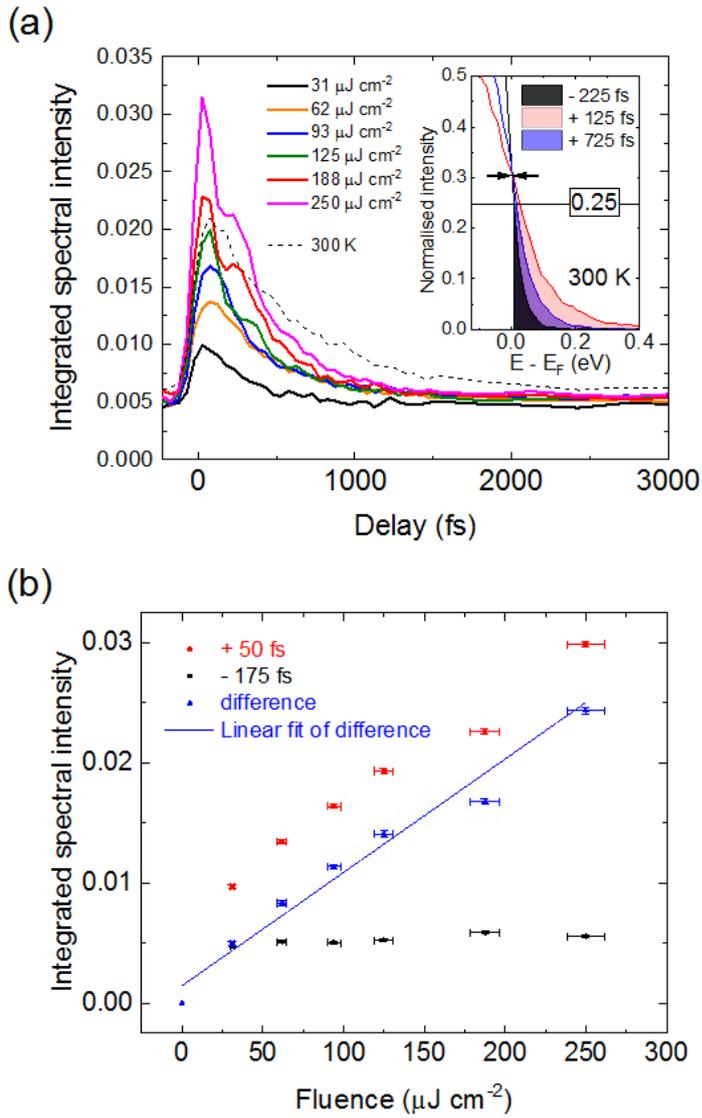

**Fig. 5.** Linearity of free carrier population above the Fermi level with pump fluence. (a) Integrated spectral intensity above the Fermi level, EF, as a function of pump-probe delay for each pump fluence in the CDW phase (80 K). The dashed line shows the normal phase (300 K) for comparison. The inset shows representative PES at 300 K and the criterion used to calculate the free-carrier population. (b) Fluence dependence of the integrated spectral intensity of the Fermi tail at -175 fs time delay (black dots), at + 50 fs time delay (red squares) and their difference (blue triangles). The linear fit gives $R^2 = 0.97377$.

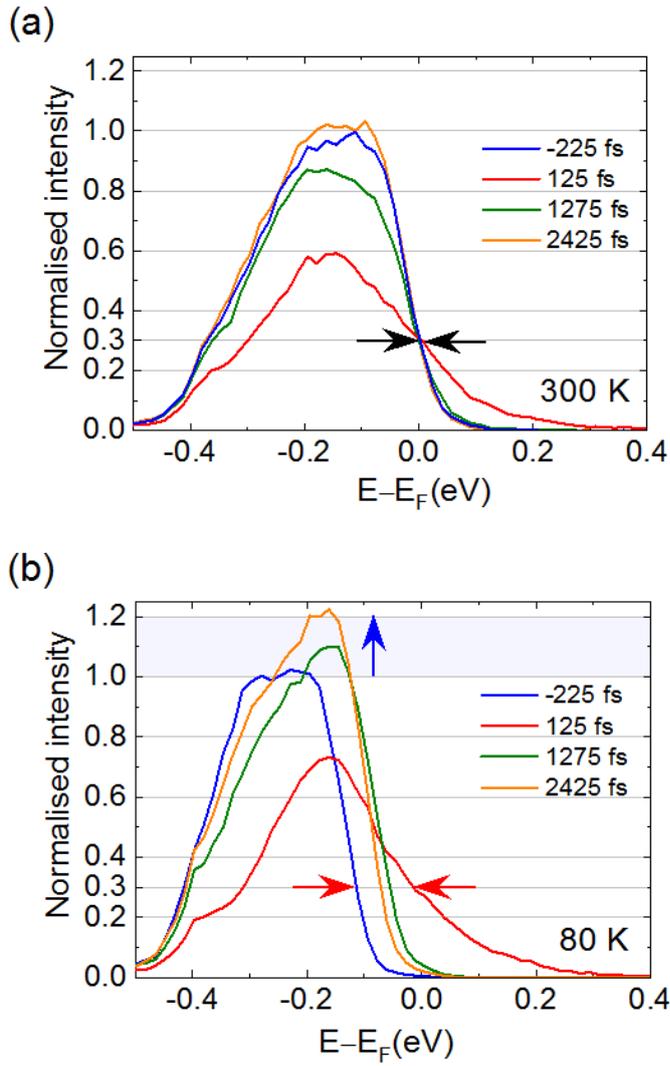

**Fig. 6.** Photo-emission spectra (PES) for various pump-probe delays. (a) Room temperature 300 K and (b) 80 K PES for fluence, F = 250 µJ cm$^{-2}$ obtained along the -14° emission angle. Curves in both panels are normalised to the intensity at - 225 fs delay. The black arrows (panel a) indicate the nodal point for all spectra in the normal phase, 300 K. At this same point in the CDW phase (b), there is a clear shift in energy at positive time delay due to VB shifting (red arrows). The blue shaded region in (b) highlights increased intensity at positive delay attributed to replica band unfolding as discussed in the main text relating to Fig. 2(d).

# References

[1] C. W. Chen, J. Choe, and E. Morosan, *Charge density waves in strongly correlated electron systems*, Rep. Prog. Phys. **79**, 084505 (2016).
[2] K. Rossnagel, *On the origin of charge-density waves in select layered transition-metal dichalcogenides*, J. Phys.-Condes. Matter **23**, 213001 (2011).
[3] J. A. Wilson, F. J. Di Salvo, and S. Mahajan, *Charge-Density Waves and superlattices in metallic layered transition-metal dichlacogenides*, Adv. Phys. **24**, 117 (1975).
[4] S. Gerber *et al.*, *Three-dimensional charge density wave order in YBa2Cu3O6.67 at high magnetic fields*, Science **350**, 949 (2015).
[5] R. Osterbacka, X. M. Jiang, C. P. An, B. Horovitz, and Z. V. Vardeny, *Photoinduced quantum interference antiresonances in pi-conjugated polymers*, Phys. Rev. Lett. **88**, 4, 226401 (2002).
[6] M. Sato, H. Fujishita, S. Sato, and S. Hoshino, *Neutron inelastic scattering and X-ray structural study of the charge density wave state in K0.3MOO3*, Journal of Physics C-Solid State Physics **18**, 2603 (1985).
[7] P. Monceau, *Electronic crystals: an experimental overview*, Adv. Phys. **61**, 325 (2012).
[8] Y. I. Joe *et al.*, *Emergence of charge density wave domain walls above the superconducting dome in 1T-TiSe2*, Nature Phys. **10**, 421 (2014).
[9] A. F. Kusmartseva, B. Sipos, H. Berger, L. Forro, and E. Tutis, *Pressure Induced Superconductivity in Pristine 1T-TiSe2*, Phys. Rev. Lett. **103**, 236401 (2009).
[10] E. Morosan *et al.*, *Superconductivity in CuxTiSe2*, Nature Phys. **2**, 544 (2006).
[11] L. J. Li, E. C. T. O'Farrell, K. P. Loh, G. Eda, B. Ozyilmaz, and A. H. C. Neto, *Controlling many-body states by the electric-field effect in a two-dimensional material*, Nature **529**, 185 (2016).
[12] J. van Wezel, P. Nahai-Williamson, and S. S. Saxena, *Exciton-phonon-driven charge density wave in TiSe2*, Phys. Rev. B **81**, 165109 (2010).
[13] M. Porer *et al.*, *Non-thermal separation of electronic and structural orders in a persisting charge density wave*, Nature Mater. **13**, 857 (2014).
[14] E. Mohr-Vorobeva *et al.*, *Nonthermal Melting of a Charge Density Wave in TiSe2*, Phys. Rev. Lett. **107**, 036403 (2011).
[15] K. Rossnagel, L. Kipp, and M. Skibowski, *Charge-density-wave phase transition in 1T-TiSe2: Excitonic insulator versus band-type Jahn-Teller mechanism*, Phys. Rev. B **65**, 235101 (2002).
[16] F. J. Di Salvo, D. E. Moncton, and J. V. Waszczak, *Electronic properties and superlattice formation in semimetal TiSe2*, Phys. Rev. B **14**, 4321 (1976).
[17] D. Jerome, T. M. Rice, and W. Kohn, *Excitonic insulator*, Physical Review **158**, 462 (1967).
[18] H. Cercellier *et al.*, *Evidence for an excitonic insulator phase in 1T-TiSe2*, Phys. Rev. Lett. **99**, 146403 (2007).
[19] G. Li, W. Z. Hu, D. Qian, D. Hsieh, M. Z. Hasan, E. Morosan, R. J. Cava, and N. L. Wang, *Semimetal-to-semimetal charge density wave transition in 1T-TiSe2*, Phys. Rev. Lett. **99**, 027404 (2007).
[20] A. Kogar *et al.*, *Signatures of exciton condensation in a transition metal dichalcogenide*, Science **358**, 1314 (2017).
[21] R. Bianco, M. Calandra, and F. Mauri, *Electronic and vibrational properties of TiSe2 in the charge-density-wave phase from first principles*, Phys. Rev. B **92**, 094107 (2015).
[22] Y. Yoshida and K. Motizuki, *Electron lattice interaction and lattice instability of 1T-TiSe2*, J. Phys. Soc. Jpn. **49**, 898 (1980).
[23] P. Chen, Y. H. Chan, X. Y. Fang, S. K. Mo, Z. Hussain, A. V. Fedorov, M. Y. Chou, and T. C. Chiang, *Hidden Order and Dimensional Crossover of the Charge Density Waves in TiSe2*, Scientific Reports **6**, 37910, 37910 (2016).

[24]	T. Pillo, J. Hayoz, H. Berger, F. Levy, L. Schlapbach, and P. Aebi, *Photoemission of bands above the Fermi level: The excitonic insulator phase transition in 1T-TiSe2*, Phys. Rev. B **61**, 16213 (2000).

[25]	M. D. Watson, O. J. Clark, F. Mazzola, I. Marković, V. Sunko, T. K. Kim, K. Rossnagel, and P. D. C. King, *Orbital- and $k_z$-Selective Hybridization of Se 4p and Ti 3d States in the Charge Density Wave Phase of TiSe$_2$*, Phys. Rev. Lett. **122**, 076404 (2019).

[26]	C. Giannetti, M. Capone, D. Fausti, M. Fabrizio, F. Parmigiani, and D. Mihailovic, *Ultrafast optical spectroscopy of strongly correlated materials and high-temperature superconductors: a non-equilibrium approach*, Adv. Phys. **65**, 58 (2016).

[27]	S. Vogelgesang *et al.*, *Phase ordering of charge density waves traced by ultrafast low-energy electron diffraction*, Nature Phys. **14**, 184 (2018).

[28]	F. Boschini *et al.*, *Collapse of superconductivity in cuprates via ultrafast quenching of phase coherence*, Nature Mater. **17**, 416 (2018).

[29]	J. Orenstein, *Ultrafast spectroscopy of Quantum Materials*, Phys. Today **65**, 44 (2012).

[30]	S. G. Han, Z. V. Vardeny, K. S. Wong, O. G. Symko, and G. Koren, *Femtosecond optical-detection of quasi-particle dynamics in high-TC YBA2CU3O7 superconducting thin films*, Phys. Rev. Lett. **65**, 2708 (1990).

[31]	S. Hellmann *et al.*, *Time-domain classification of charge-density-wave insulators*, Nature Comm. **3**, 1069 (2012).

[32]	T. Rohwer *et al.*, *Collapse of long-range charge order tracked by time-resolved photoemission at high momenta*, Nature **471**, 490 (2011).

[33]	C. Monney *et al.*, *Revealing the role of electrons and phonons in the ultrafast recovery of charge density wave correlations in 1T-TiSe2*, Phys. Rev. B **94**, 165165 (2016).

[34]	c. p. o. See Supplemental Material at [URL will be inserted by publisher] for [Electronic trasnport, laser heating, simulations and Raman].

[35]	F. Boschini *et al.*, *An innovative Yb-based ultrafast deep ultraviolet source for time-resolved photoemission experiments*, Rev. Scien. Instrum. **85**, 123903 (2014).

[36]	S. Mathias *et al.*, *Self-amplified photo-induced gap quenching in a correlated electron material*, Nature Comm. **7**, 12902 (2016).

[37]	G. Rohde *et al.*, *Tracking the relaxation pathway of photo-excited electrons in 1T-TiSe2*, Eur. Phys. J.-Spec. Top. **222**, 997 (2013).

[38]	T. Limmer, J. Feldmann, and E. Da Como, *Carrier Lifetime in Exfoliated Few-Layer Graphene Determined from Intersubband Optical Transitions*, Phys. Rev. Lett. **110**, 217406 (2013).

[39]	D. Bugini *et al.*, *Ultrafast spin-polarized electron dynamics in the unoccupied topological surface state of Bi 2 Se 3*, Journal of Physics: Condensed Matter **29**, 30LT01 (2017).

[40]	J. H. Buss *et al.*, *A setup for extreme-ultraviolet ultrafast angle-resolved photoelectron spectroscopy at 50-kHz repetition rate*, Rev. Scien. Instrum. **90**, 11, 023105 (2019).

[41]	L. Perfetti *et al.*, *Time evolution of the electronic structure of 1T-TaS2 through the insulator-metal transition*, Phys. Rev. Lett. **97**, 067402 (2006).

[42]	J. A. Holy, K. C. Woo, M. V. Klein, and F. C. Brown, *Raman and infrared studies of superlattice formation in TiSe2*, Phys. Rev. B **16**, 3628 (1977).

[43]	T. Huber *et al.*, *Coherent Structural Dynamics of a Prototypical Charge-Density-Wave-to-Metal Transition*, Phys. Rev. Lett. **113**, 026401 (2014).

[44]	L. Rettig *et al.*, *Persistent order due to transiently enhanced nesting in an electronically excited charge density wave*, Nature Comm. **7**, 10459 (2016).

[45]	A. Rothwarf and B. N. Taylor, *Measurement of recombination lifetimes in superconductors*, Phys. Rev. Lett. **19**, 27 (1967).

[46]	V. V. Kabanov, J. Demsar, and D. Mihailovic, *Carrier-relaxation dynamics in intragap states: The case of the superconductor YBa$_2$Cu$_3$O$_7$ and the charge-density-wave semiconductor K$_{0.3}$MoO$_3$*, Phys. Rev. B **61**, 1477 (2000).

[47]     P. Beaud *et al.*, *A time-dependent order parameter for ultrafast photoinduced phase transitions*, Nature Mater. **13**, 923 (2014).
[48]     F. Weber *et al.*, *Electron-Phonon Coupling and the Soft Phonon Mode in TiSe2*, Phys. Rev. Lett. **107**, 5, 266401 (2011).
[49]     X. Shi *et al.*, *Ultrafast electron calorimetry uncovers a new long-lived metastable state in 1T-TaSe$_2$ mediated by mode-selective electron-phonon coupling*, Science Advances **5**, eaav4449 (2019).
[50]     C. Sohrt, A. Stange, M. Bauer, and K. Rossnagel, *How fast can a Peierls–Mott insulator be melted?*, Faraday Discussions **171**, 243 (2014).
[51]     S. Sugai, *Lattice-vibrations in the charge density-wave states of layered transition-metal dichalcogenides*, Phys. Stat. Sol. B-Basic Research **129**, 13 (1985).
[52]     S. Wall, D. Wegkamp, L. Foglia, K. Appavoo, J. Nag, R. F. Haglund, J. Stahler, and M. Wolf, *Ultrafast changes in lattice symmetry probed by coherent phonons*, Nature Comm. **3**, 6, 721 (2012).
[53]     C. Monney *et al.*, *Probing the exciton condensate phase in 1T-TiSe2 with photoemission*, New J. Phys. **12**, 32, 125019 (2010).
[54]     T. E. Kidd, T. Miller, M. Y. Chou, and T. C. Chiang, *Electron-Hole Coupling and the Charge Density Wave Transition in TiSe$_2$*, Phys. Rev. Lett. **88**, 226402 (2002).